\documentclass[11pt]{article}
\usepackage{amsfonts}
\usepackage{amsmath}
\usepackage{amssymb} 
\usepackage[compress]{cite}
\usepackage{bm} 
\usepackage{color}
\usepackage{comment}
\includecomment{comment}
\usepackage{dcolumn}
\usepackage{enumerate}
\usepackage[margin=1in]{geometry}                
\usepackage{graphicx}
\usepackage{epstopdf}   
\usepackage{mathrsfs}
\usepackage{mathtools} 
\usepackage{paralist}
\usepackage[parfill]{parskip}    
\usepackage{textcomp}
\usepackage{titlesec}
\usepackage{varioref}
\usepackage{hyperref} 
\allowdisplaybreaks
\usepackage{stackengine}
\def\dcirc#1{\stackinset{c}{}{t}{-3.5pt}%
  {$\mkern2.5mu\scriptscriptstyle\circ\mkern-1mu\circ$}{$#1$}}
\usepackage{accents}
\hypersetup{
    pdfstartview={FitH},    
    colorlinks=true,       
    linkcolor=blue,          
    citecolor=blue,        
    filecolor=blue,      
    urlcolor=magenta           
}
\DeclareFontFamily{OT1}{pzc}{}
\DeclareFontShape{OT1}{pzc}{m}{it}%
             {<-> s * [0.900] pzcmi7t}{}
\DeclareMathAlphabet{\mathscr}{OT1}{pzc}%
                                 {m}{it}
\labelformat{section}{Section #1} 
\labelformat{subsection}{Section #1} 
\labelformat{equation}{Eq.~(#1)} 
\labelformat{figure}{Fig.~#1} 
\labelformat{subfigure}{Fig.~\thefigure#1} 
\labelformat{table}{Tab.~#1} 
\titleclass{\subsubsubsection}{straight}[\subsection]

\newcounter{subsubsubsection}[subsubsection]
\renewcommand\thesubsubsubsection{\thesubsubsection.\arabic{subsubsubsection}}

\titleformat{\subsubsubsection}
{\normalfont\normalsize\bfseries}{\thesubsubsubsection}{1em}{}
\titlespacing*{\subsubsubsection}
{0pt}{3.25ex plus 1ex minus .2ex}{1.5ex plus .2ex}
\makeatletter
\renewcommand\paragraph{\@startsection{paragraph}{5}{\z@}%
	{3.25ex \@plus1ex \@minus.2ex}%
	{-1em}%
	{\normalfont\normalsize\bfseries}}
\renewcommand\subparagraph{\@startsection{subparagraph}{6}{\parindent}%
	{3.25ex \@plus1ex \@minus .2ex}%
	{-1em}%
	{\normalfont\normalsize\bfseries}}
\def\toclevel@subsubsubsection{4}
\def\toclevel@paragraph{5}
\def\toclevel@paragraph{6}
\def\l@subsubsubsection{\@dottedtocline{4}{7em}{4em}}
\def\l@paragraph{\@dottedtocline{5}{10em}{5em}}
\def\l@subparagraph{\@dottedtocline{6}{14em}{6em}}
\makeatother
\def\f{\frac}

\def\l{\left}
\def\r{\right}

\def\Mpl{M_{_{\mathrm{Pl}}}}

\newcommand{\calO}{{\cal O}}
\newcommand{\calP}{{\cal P}}
\newcommand{\calR}{{\cal R}}

\title{
Mimicking two field inflationary  features with a single field }
\author{Anvy Moly Tom\footnote{anvymolytom@gmail.com}$~^{1}$ and Rathul Nath Raveendran \footnote{rathulnath.r@gmail.com}$~^{2}$\\
{\small{$^{1}$ Department of Physics}}\\ 
{\small{Devaswom Board College Thalayolaparambu, Kottayam~686605, India}}\\
{\small{$^{2}$ School of Physical Sciences}}\\ 
{\small{Indian Association 
for the Cultivation of Science, Kolkata~700032, India}}
}

\begin{document}

\maketitle
\begin{abstract}
It is well known that damped superimposed oscillations at large scales in the primordial power spectrum can be generated in both single field and two field models. In single field inflationary models, these features typically arise due to deviations from the slow roll regime. On the other hand, in two field models, these features are generated due to a turn in the background trajectory in the field space. In this work, we demonstrate that both single field and two field models can produce identical features at large scales in the primordial power spectrum. To achieve this, we utilize the generalized slow roll approximation and successfully reconstruct single field models based on the featured power spectrum typically generated in two field models. To validate our methodology, we numerically calculate the power spectrum from the reconstructed potential and find a remarkable agreement with the power spectrum which is obtained from the two field model.
\end{abstract}

\section{Introduction}
\label{sec:intro}


WMAP and Planck have made precise observations of anisotropies in the Cosmic Microwave Background (CMB), which suggest a primordial scalar power spectrum that is almost scale-independent and predominantly adiabatic~\cite{WMAP:2003elm,WMAP:2003ivt,Planck:2018jri}. This characteristic perturbation can be generated by the inflationary scenario, which is driven either by scalar fields~\cite{Guth:1980zm,Linde:1981mu} or purely geometrically in modified gravity~\cite{Starobinsky:1980te}. Slow roll inflation models can produce power spectra consistent with cosmological data. Although a nearly scale-invariant primordial power spectrum is consistent with observations, there have been repeated attempts to detect features in the power spectrum~\cite{Starobinsky:1992ts,Covi:2006ci,Miranda:2013wxa,Hazra:2016fkm,Adams:2001vc}. These features have been found to improve the fit to the CMB data, which is significant because confirming them through future observations would significantly constrain the space of viable models.


The inflationary epoch is usually
driven with the aid of one or more scalar fields. To generate oscillations in single field models, a brief deviation from slow roll inflation is necessary. This deviation is typically accomplished by introducing a small step or bump in the potential~\cite{Starobinsky:1992ts, Miranda:2013wxa,Adams:2001vc,Hazra:2010ve}. However, because the trajectory is an attractor, slow roll inflation is eventually restored after these temporary departures.

When considering two field models, it is recognized that the heavy degrees of freedom in the model, which are relatively heavy compared to the inflationary scale, can produce curvature power spectrum features~\cite{Achucarro:2010da,Cespedes:2012hu,Chen:2012ja, Gao:2012uq,Braglia:2020fms}. Notably, some of these features generated by the models were found to improve the fit to the CMB data. It is suggested that if these features are present, they could offer new ways to test the multi-field paradigm and provide compelling evidence for the existence of additional degrees of freedom during inflation.


However, this work demonstrates that the features created in two field models can also be produced by single field models. To illustrate this, we utilize the general slow roll approximation (GSR)~\cite{Stewart:2001cd,Choe:2004zg,Joy:2005ep,Joy:2005pe,Choi:2021key} and present a technique for directly reconstructing the inflationary potential in single field models based on the power spectrum generated in two field models.

 This article is structured as follows: In \ref{sec:twofield}, we provide an overview of two field models, including the evolution of perturbations. We also construct models that exhibit features in the power spectrum by introducing relevant background functions. Moving on to \ref{sec:eom-perturbations}, we delve into the evolution of perturbations within single field models. \ref{sec:Reconstruction-formalism} is dedicated to reviewing the Generalized Slow Roll (GSR) formalism used for reconstructing the inflationary potential. Subsequently, in \ref{sec:reconstrvtion-feature}, we utilize the GSR formalism to reconstruct the inflationary potential of single field models based on the distinctive power spectrum features obtained from two field models. Finally, we provide a brief conclusion in \ref{sec:conc}.

At this stage of our discussion, let us clarify the conventions and notations that will be adopted in this work. We will utilize natural units, setting $\hbar$ and $c$ to 1, while defining the Planck mass as $\Mpl=(8 \pi G)^{-1/2}$. In line with common practice, differentiation with respect to the cosmic time $t$ will be denoted by an overdot, while differentiation with respect to the conformal time $\eta$ will be denoted by an overprime.

\section{Two field models}\label{sec:twofield}
In this section, we shall discuss the evolution and power spectra of perturbations in two field models. The focus will be on the power spectra of specific two field models that exhibit features.
\subsection{Evolution of perturbations}

In the context of inflationary models that involve two fields, it is well-known that isocurvature perturbations can arise in addition to curvature perturbations. At the linear perturbation theory level, we can construct a Hamiltonian that is associated with the Fourier components of both the curvature and entropic perturbations as \cite{Langlois:2008qf,Langlois:2008mn},
\begin{equation} 
H_2=\frac{1}{2} \left(p_\sigma^2 + p_s^2\right) +p_\sigma \,\frac{z'}{z} \, v_\sigma+p_\sigma \, \xi \, v_s+p_s \, \frac{a'}{a} \, v_s +\frac{k^2}{2}\, v_\sigma^2 + \frac{1}{2}\left(k^2  + \mu_{s}^2\,a^2+\xi^2\right) v_s^2~,
\label{eq:H-tfm}
\end{equation}
where, $v_{\sigma}$ and $v_{s}$ represent the curvature and entropic perturbations, respectively, and their conjugate momenta are denoted by $p_{\sigma}$ and $p_s$. The couplings between $p_{\sigma}$ with $v_{\sigma}$ and $v_{s}$ are denoted by $z(\eta)$ and $\xi(\eta)$, respectively, whereas the coupling between $p_{s}$ and $v_{s}$ is dependent on the scale factor $a(\eta)$. Additionally, $k$ corresponds to the magnitude of the wave vector, and $\mu_{s}$ is a time-dependent mass scale that is associated with the entropic perturbation $v_{s}$. It should be noted that specific models require the time-varying functions $\xi(\eta)$, $\mu_s(\eta)$, $z'/z$, and the scale factor $a(\eta)$ to be determined. From the above expression, Hamilton's equations can be obtained as, 
\begin{eqnarray}
\label{eq:cl-H-eqs-tfm}
p_\sigma &=& v_{\sigma}'-\frac{z'}{z}\, v_{\sigma}-\xi\, v_s~,
\qquad 
p_s = v_{s}'- \frac{a'}{a}\, v_s~,\\
p_{\sigma}'&= &- \frac{z'}{z}\, p_{\sigma}-k^2\, v_{\sigma}~,
\qquad
p_{s}'= - \frac{a'}{a}\, p_{s}-\xi\, p_{\sigma}
-\left(k^2+\mu_{s}^2\,a^2 +\xi^2\right)\, v_{s}~.
\end{eqnarray}
From the above equations, the equations of motion governing the variables $v_\sigma $ and $v_s$ can be obtained to be
\begin{subequations}
\label{eq:ms}
	\begin{eqnarray} 
		v_{\sigma}'' 
		+\l(k ^2 - \f{z''}{z}\r)\, v_{\sigma}
		&= &\frac{1}{z}\, \l(z\, \xi\, v_s\r)'~,\\
		v_s'' + \l(k^2- \f{a''}{a}+a^2\, \mu_{s}^2\r)\, v_{s}
		&=&-z\, \xi\, \l(\f{v_\sigma }{z}\r)'~.
	\end{eqnarray}
\end{subequations}

It is well known that the perturbations under consideration originated as quantum fluctuations. To quantize the perturbations, we can promote the relevant variables to quantum operators as
\begin{eqnarray}
\hat{v}_{\sigma}&=& f_{\sigma} \hat{a} + f_{\sigma}^\star \hat{a}^{\dagger}+g_{\sigma} \hat{b} + g_{\sigma}^\star \hat{b}^{\dagger}~, \\
\hat{v}_{s}&=& f_{s} \hat{a} + f_{s}^\star \hat{a}^{\dagger}+g_{s} \hat{b} + g_{s}^\star \hat{b}^{\dagger}~,
\end{eqnarray}
where $f_{\sigma, s}$ and $g_{\sigma, s}$ are the solutions of \ref{eq:ms}, and $(\hat{a}, \hat{b})$ and $(\hat{a}^\dagger, \hat{b}^\dagger)$ are the annihilation and creation operators. The vacuum states are defined as
\begin{equation}
\hat{a} \rvert 0 \rangle = \hat{b} \rvert 0 \rangle = 0~.
\end{equation}
When the modes are well inside the Hubble radius, the equations of motion governing the variables ($f_\sigma, f_s$) and ($g_\sigma, g_s$) are decoupled, and we can set the initial conditions using the Minkowski-like vacuum as
\begin{eqnarray}
f_{\sigma}(\eta)&=& g_{s}(\eta)=\frac{e^{-i k \eta}}{\sqrt{2 k}}~,\\
f_{s}(\eta)&=& g_{\sigma}(\eta)=0~.
\end{eqnarray}
The scalar power spectra can be expressed as~\cite{Lalak:2007vi, Garcia-Bellido:1995hsq}
\begin{eqnarray}
P_\calR &=& \frac{k^3}{2 \pi^2} \frac{\vert f_{\sigma}\vert^2 + \vert g_{\sigma}\vert^2}{z^2}~, \\
P_S &=& \frac{k^3}{2 \pi^2} \frac{\vert f_s\vert^2 + \vert g_s\vert^2}{z^2}~,
\label{eq:PR-PS-def}
\end{eqnarray}
where $z(\eta)$ is the coupling between $p_\sigma$ and $v_\sigma$, and $k$ is the magnitude of the wave vector. These equations can be solved analytically, and by imposing the Bunch-Davies initial conditions at early times, we can evaluate the scalar power spectra at later times, closer to the end of inflation.

\subsection{Models with featured power spectrum}
\label{sec:2-f-feature}
Our next task is to build two field models capable of producing a power spectrum with distinct features that align with CMB observations. As mentioned earlier, a turn in the background during the exit of cosmologically relevant scales through the Hubble horizon can result in specific features in the curvature perturbation power spectrum. The appropriate selection of background functions can make these features consistent with CMB observations. In particular, we must determine the time evolution of the functions $z$, $\xi$, and $\mu_{s}$.

For convenience, the following slow roll parameters are often introduced, 
\begin{subequations}\label{eq:def-epsilons}
		\begin{align}
    \epsilon_1=& - \frac{{\rm d\, log}H}{{\rm d} N}~,
    \\
    \epsilon_{n+1}=& \frac{{\rm d\, log}\epsilon_n}{{\rm d} N}~,
		\end{align}
\end{subequations}
where, $H$ is the Hubble parameter and $N$ corresponds to the e-folding parameter, defined as
\begin{equation}
d N=aH\,d\eta~.
\end{equation}
To simplify notation, we express $(z'/z)$ in terms of the second slow roll parameter $\epsilon_{2}$ as
\begin{equation}\label{eq:zpz-in-es}
\frac{z'}{z}=a\, H\left( 1+\frac{\epsilon_2}{2}\right)~,
\end{equation}
where, as mentioned before, $H$ is the Hubble parameter and the slow roll parameters are defined in \ref{eq:def-epsilons}. 

Note that the e-folding parameter is a monotonic function of the conformal time in the inflationary epoch and hence all the time derivative operators can be expressed by derivatives with respect to the e-folding parameter $N$, as in \ref{eq:def-epsilons}. In accordance with \cite{Raveendran:2023dst}, we assume a constant second slow roll parameter. Then the first slow roll parameter and the Hubble parameter can be obtained using \ref{eq:def-epsilons} as
\begin{subequations}\label{eq:e1-H}
\begin{eqnarray}
\epsilon_1(N)&=&\epsilon_1(N_p) \, {\rm exp}\left[\epsilon_2\left( N-N_p\right)\right]~,
\\
H(N)&=&H(N_p)\,{\rm exp}\left[\frac{\epsilon_1(N) - \epsilon_1(N_p)}{\epsilon_2}\right]~,
\end{eqnarray}
\end{subequations}
where $N_p$ is the time (expressed in terms of e-folding) at which the pivot scale $k=0.05\,{\rm Mpc^{-1}}$ exits the Hubble horizon. 
Furthermore, we adopt a simplified form for the coupling parameter $\xi$ based on  \cite{Achucarro:2010da} as
\begin{equation}\label{eq:xibaH}
\frac{\xi}{a H} = \xi_{m} \, {\rm sech}^2\left[\frac{(N-N_0)}{\Delta}\right]~,
\end{equation}
where, $\xi_{m}$, $N_0$, and $\Delta$ are constants. 
We choose the value of $N_0$ for the coupling function $\xi$ such that the change in the trajectory occurs when cosmologically relevant scales exit the Hubble horizon, which is known to induce features in the curvature perturbations. Moreover, the e-folding parameter $N_{0}$ indicates the time when the coupling between curvature and entropic perturbation becomes significant. Similarly, the mass $\mu_s$ associated with the entropic perturbation can be expressed as
\begin{equation}
\mu_s^2= M^2-\left(\frac{\xi}{2a}\right)^2=M^{2}\left[1-\left(\frac{H \xi_{m}}{2M}\right)^{2}{\rm sech}^4\left[\frac{(N-N_0)}{\Delta}\right] \right]~, 
\end{equation}
where, the quantity $M$ is another constant.

\begin{figure}[h!]
\centering
\includegraphics[width=0.75\linewidth]{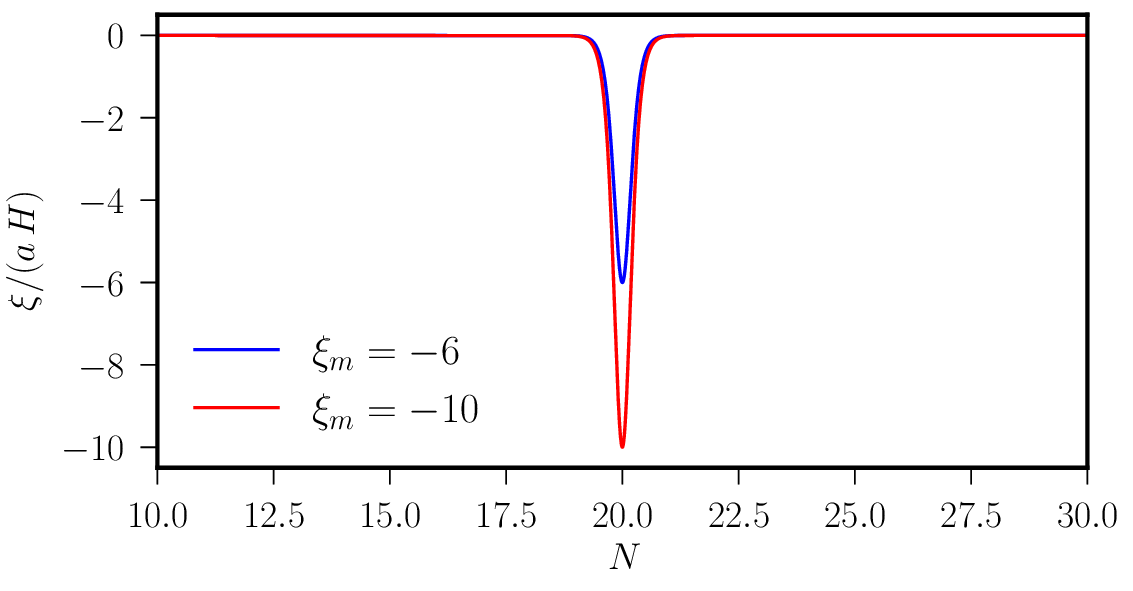}
\includegraphics[width=0.75\linewidth]{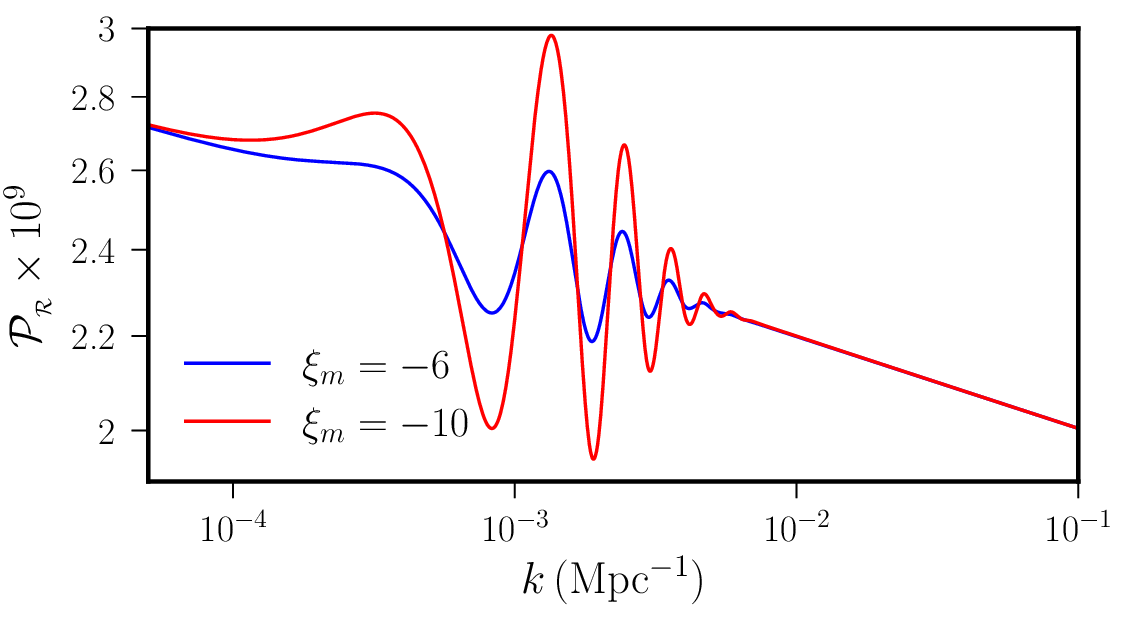}
\caption{The evolution of coupling parameter $\xi$ and the associated power spectra have been plotted for different values of the amplitude $\xi_m$. A non-zero value of $\xi$ indicates a change in the background trajectory. Notably, the peak of $\xi$ at $N_0=20$ introduces distinctive features in the curvature power spectrum. These plots are created using the following parameter values: $\epsilon_2=0.036$, $\epsilon_1(N_p) = 1.875\times10^{-3}$, and $H(N_p) =1.724 \times 10^{-5} M_{{\mathrm{Pl}}}$. Consequently, the scalar spectral index is determined as $n_{\rm s} =0.96$, and the amplitude of the scalar power spectrum is $ A_{\rm s} = 2\times 10^{-9}$.}
\label{fig:xibaH-N0-35}
\end{figure}

  Our previous focus has been on determining the background functions necessary for generating a power spectrum with distinct features in the context of two field inflation. Now, our task is to solve equations \ref{eq:ms} and obtain the power spectrum. By making reasonable parameter choices, we can obtain scalar power spectra that exhibit these features. \ref{fig:xibaH-N0-35} illustrates the evolution of the quantity $\xi$ with the e-folding parameter for various values of $\xi_{m}$, along with the corresponding power spectra. Notably, the power spectrum is influenced by the amplitude $\xi_{\rm m}$ of the coupling function. Different values of $\xi_{\rm m}$ lead to significantly diverse power spectrum magnitudes, while preserving the overall structure. This observation indicates that any abrupt deviation from the background trajectory can induce oscillatory features in the power spectrum.


\section{Evolution of perturbations in single field models}
\label{sec:eom-perturbations}

In the last section, we have discussed the equations governing the perturbations in the case of two field models. Now, shifting our focus to single field inflationary models, we can derive the equation of motion for the Mukhanov-Sasaki variable, denoted as $v$, from \ref{eq:ms}. This expression can be written as
\begin{equation}
v''+\l(k^2-\f{z''}{z}\r)v=0\, .\label{eq:de-v}
\end{equation}
Similarly, as in the case of two field models, the power spectrum is defined as
\begin{equation}
P_\calR = \f{k^3}{2 \pi^2}\f{|f|^2}{z^2} \,,
\end{equation}
where $f$ is the mode function associated with the operator $\hat{v}$ and it obeys \ref{eq:de-v}.
Usually, as in the case of two field models, Bunch-Davies initial conditions are imposed on the 
Mukhanov-Sasaki variable~$v$ when the modes are well inside the Hubble 
radius, i.e. when $k \gg (a\,H)$ or, more precisely, when $k \gg
\sqrt{z''/z}$. We choose the initial condition for the quantity~$f$ as

\begin{equation}
f(\eta_i) = \f{1}{\sqrt{2\,k}}\,\mathrm{e}^{-i\,k\,\eta_i},
\end{equation}
where $\eta_i$ is the time when the modes are in the sub-Hubble domain.
\section{Reconstructing single field models using GSR}
\label{sec:Reconstruction-formalism}
In this section, we provide a brief overview of the general slow roll approximation scheme utilized for reconstructing the inflationary potential~\cite{Stewart:2001cd, Choe:2004zg,Choi:2021key, Joy:2005ep,Joy:2005pe}. The first step is to rewrite \ref{eq:de-v} as
\begin{equation}
\label{eq:modeeq1}
\frac{d^2 v}{ds^2} + \bigg( k^2 - \frac{1}{z} \frac{d^2z}{ds^2} \bigg) v = 0 \, ,
\end{equation}
where, we have defined the positive conformal time $s$ as
\begin{equation}
\label{eq:xi-def}
s \equiv - \int \frac{dt}{a} = \frac{1}{aH} \big[ 1 + \calO(\epsilon) \big] \, .
\end{equation}

Next, we can express \ref{eq:modeeq1} in a form that clearly distinguishes between the contribution of the exactly scale-invariant power spectrum and deviations from this spectrum.
~To achieve this, we introduce the variables $y \equiv \sqrt{2k}\varphi$ and $x \equiv ks$, and rescale $z$ as
\begin{equation}
\label{eq:GSRfunction}
f(\log s) \equiv \frac{2\pi x}{k}z \, .
\end{equation}
With these transformations, \ref{eq:modeeq1} transforms into
\begin{equation}
\label{eq:modeeq2}
\frac{d^2y}{dx^2} + \bigg( 1 - \frac{2}{x^2} \bigg) y = \frac{1}{x^2} g(\log s) y \, ,
\end{equation}
where
\begin{equation}
g \equiv \frac{\dcirc{f}-3\accentset{\circ}{f}}{f}~,
\end{equation}
The circle over the variable here and throughout are derivatives with respect
to $d\log s$. Note that, in \ref{eq:modeeq2}, the left-hand side represents the mode function equation in the perfect de Sitter background. The corresponding power spectrum of the solution can be written as
\begin{equation}
\label{eq:GSRpower}
\calP_\calR(k) = \lim_{x\to 0} \bigg| \frac{xy}{f} \bigg|^2 \,~.
\end{equation}
We emphasize that the function $g$ on the right-hand side of \ref{eq:modeeq2} accounts for possible deviations from scale invariance in the power spectrum.

To the leading order of GSR, we can solve for $y(x)$ using the Green's function method and from \ref{eq:GSRpower} the power spectrum can be expressed as~\cite{Stewart:2001cd,Choe:2004zg,Dvorkin:2009ne}
\begin{equation}
\label{eq:GSRspectrum}
\log\calP_\calR(k)
=
\int_0^\infty \frac{ds}{s} \big[ -\accentset{\circ}{W}(ks) \big]
\bigg[ \log \bigg( \frac{1}{f^2} \bigg) + \frac{2}{3} \frac{\accentset{\circ}{f}}{f}\bigg]
\,~,
\end{equation}
where $W(x)$ represents the window function given by
\begin{equation}
\label{eq:window}
W(x) = \frac{3\sin(2x)}{2x^3} - \frac{3\cos(2x)}{x^2} - \frac{3\sin(2x)}{2x} - 1 \,~.
\end{equation}
%
In the standard slow roll approximation, as evident from \ref{eq:GSRspectrum}, $\accentset{\circ}{f}/f$ varies slowly and the power spectrum is approximately given by the well-known result $\mathcal{P}_\mathcal{R} \approx [H^2/(2\pi \dot{\phi})]^2$. However, within the GSR framework, $\accentset{\circ}{f}/f$ may vary rapidly while remaining small, resulting in prominent features in the power spectrum. Furthermore, in GSR, the evolution of each mode throughout the entire inflationary epoch is automatically accounted for due to the integration range, whereas the standard slow roll approximation only considers effects around the moment of horizon crossing, $k=aH$.

It is interesting to note that, \ref{eq:GSRspectrum} represents an integral transformation between $\calP_\calR(k)$ and $f(\log s)$ utilizing the window function \ref{eq:window}. Interestingly, it has been known that this expression can be inverted and can be used for obtaining the function $f(\log s)$ from a given power spectrum as ~\cite{Joy:2005ep,Joy:2005pe}
\begin{equation}
\label{eq:inverse}
\log \bigg( \frac{1}{f^2} \bigg) = \int_0^\infty \frac{dk}{k} m(k s)\log\calP_\calR(k)~,
\end{equation}
where
\begin{equation}
m(x) = \frac{2}{\pi} \bigg[ \frac{1}{x} - \frac{\cos(2x)}{x} - \sin(2x) \bigg] \, .
\end{equation}

Once the function $f(\log s)$ is obtained, it becomes possible to determine the Hubble parameter by solving the Friedmann equation
\begin{equation}
\label{eq:Heq}
\dot{H} = - \frac{\dot\phi^2}{2\Mpl^2} = - \frac{H^2f^2}{2(2\pi)^2\Mpl^2 a^2 s^2} \,~,
\end{equation}
where in the second equality we have used \ref{eq:GSRfunction} to eliminate $\dot\phi$. In the limit, $\epsilon \ll 1$, using $ s \approx 1/(aH)$, we find the following differential equation for $H$:
\begin{equation}
H^{-3} \frac{dH}{d s} = \frac{1}{2(2\pi)^2\Mpl^2} \frac{f^2}{ s} \, .
\end{equation}
Simultaneously, by employing the Friedmann equations~\ref{eq:Heq} and \ref{eq:GSRfunction}, it is straightforward to express the potential as a function of $\log s$
\begin{equation}
\label{eq:potential}
V = 3\Mpl^2H^2 - \frac{1}{2}\dot\phi^2
= 3\Mpl^2H^2 \bigg[ 1 - \frac{f^2H^2}{6(2\pi)^2\Mpl^2} \bigg]
\, .
\end{equation}
Finally, by utilizing~\ref{eq:GSRfunction}, we can derive the differential equation for $\phi$ as
\begin{equation}
\label{eq:phieq}
\frac{d\phi}{d\log s} = -\frac{fH}{2\pi} \, .
\end{equation}
By eliminating $s$ in \ref{eq:potential} using $\phi(\log s)$, it becomes possible to reconstruct the potential as a function of the inflaton field, $V(\phi)$.



 \color{black}

\section{Reconstructing single field potential from featured power spectrum}\label{sec:reconstrvtion-feature}
In the previous section, we have outlined the methodology for reconstructing the potential when the power spectrum is known. Now, we will apply this formalism to power spectra obtained in two field models, as described in ~\ref{sec:2-f-feature}.

\begin{figure}[h!]
\centering
\includegraphics[width=0.49\linewidth]{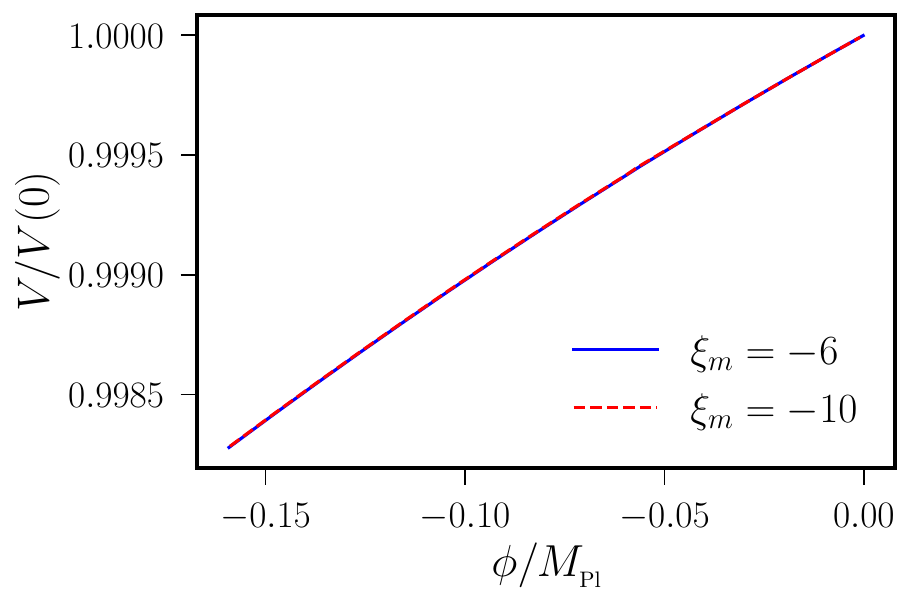}
\includegraphics[width=0.49\linewidth]{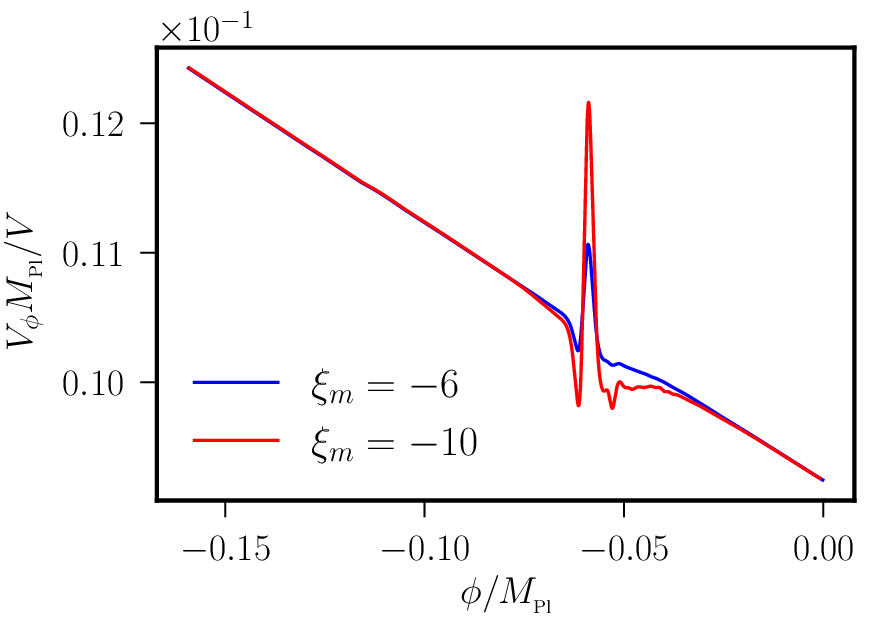}
\includegraphics[width=0.85\linewidth]{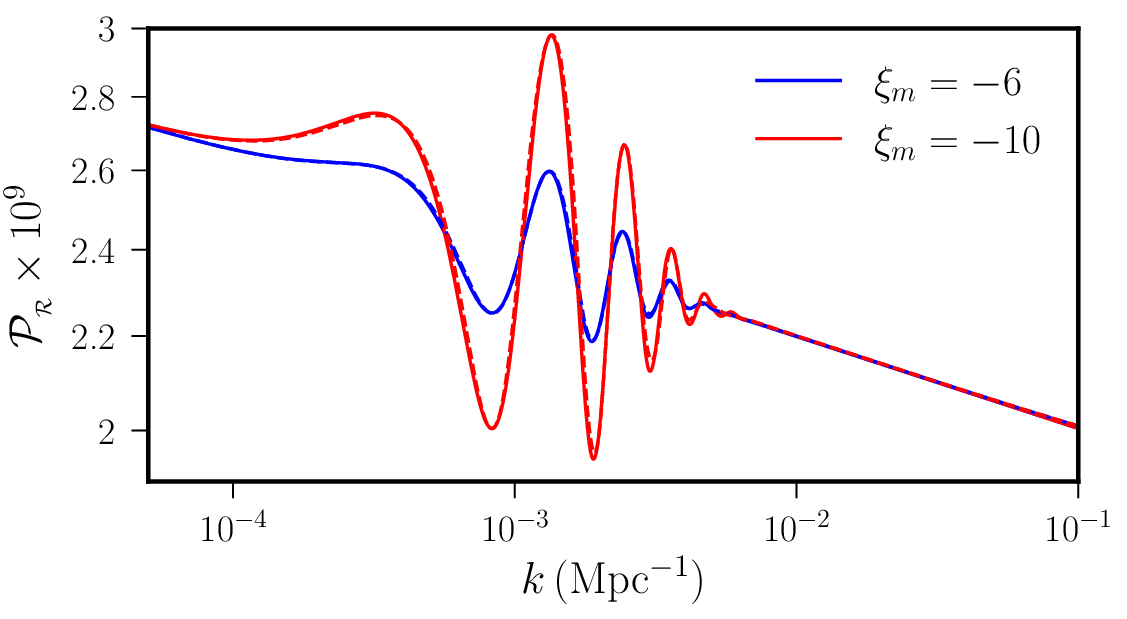}
\caption{The figures show the reconstructed potentials, their derivatives, and power spectra calculated from the reconstructed potential. The power spectra calculated numerically from the reconstructed potentials are represented by the dashed lines in the bottom figure. In order to make a comparison, the power spectra obtained from two field models are also displayed as solid lines. As apparent from the figure, the power spectra derived from the reconstructed models exhibit excellent agreement with the input power spectra, which correspond to the spectra obtained in two field models.}
\label{fig:V-Ps}
\end{figure}

\ref{fig:V-Ps} illustrates the reconstructed potential derived from the featured power spectrum using the formalism discussed in~\ref{sec:Reconstruction-formalism}. To demonstrate the changes in the potential responsible for the observed features, we have included the slope of the potential as well. Furthermore, as a consistency check, we have numerically calculated the power spectrum from the reconstructed potential. Remarkably, the resulting power spectrum exhibits excellent agreement with the input power spectrum, which corresponds to the spectrum obtained in two field models and is also displayed in the figure.

As previously mentioned, in single field models, features arise from deviations from the slow roll trajectory, primarily caused by variations in the slope of the potential, as illustrated in \ref{fig:V-Ps}. In contrast, in two field models, features emerge from the interaction between curvature and isocurvature perturbations. Remarkably, our results suggest that these identical features can be generated by entirely distinct mechanisms during inflation.

\section{Conclusions}
\label{sec:conc}
This study demonstrates that the distinctive features which have the form of damped superimposed oscillations in the primordial power spectrum generated by two field models can also be replicated by single field models. Specifically, through the use of the generalized slow roll approximation (GSR), we successfully reconstruct a single field potential that produces a power spectrum with similar features as obtained from two field models.
To verify the consistency of our approach, we numerically calculate the power spectrum from the reconstructed potential. The results exhibit a high level of agreement with the input power spectrum, as depicted in~\ref{fig:V-Ps}.
Therefore, it is possible to conclude that distinguishing between single field and two field models based solely on the constraints derived from the large-scale features in the power spectrum is not possible. In other words, when models are constrained using these features in the power spectrum, these models are essentially equivalent. {\color{black} It is important to note that a similar conclusion was reached in~\cite{Braglia:2022ftm}, which compared various feature models. However, our approach is distinct as it relies on reconstruction using the GSR approximation method from a specific featured power spectrum}. Nevertheless, it is possible to differentiate between single field and two field models by examining other two-point correlation functions, such as momentum correlation and momentum-curvature perturbation correlations, as discussed in \cite{Raveendran:2023dst}.

There are also several directions to explore in the future. Our findings have revealed that in terms of constraining models using large-scale features in the power spectrum, single field and two field models are essentially indistinguishable. Notably, it is recognized that the amplification in the power spectrum, which leads to black hole formation, can arise from both single field~\cite{Ivanov:1994pa,Garcia-Bellido:2017mdw, Ragavendra:2020sop,Ragavendra:2023ret} and two field models~\cite{Clesse:2015wea,palma2020seeding, Braglia:2020eai,Raveendran:2023dst,Choi:2021yxz}. In light of this, it would be fascinating to investigate the possibility of constructing distinct single field and two field models that yield identical power spectrum shapes within the context of primordial black hole formation.

\subsection*{Acknowledgements}

We thank Matteo Braglia for bringing the important paper~\cite{Braglia:2022ftm} to our attention. The authors wish to thank Krishnamohan Parattu for his comments on the manuscript. RNR is supported by post-doctoral fellowship from the Indian Association for the Cultivation of Science, Kolkata, India. AMT is supported by SR/FST/COLLEGE-420/2018.

During the preparation of this work the authors used chat gpt in order to check grammar. After using this tool, the authors reviewed and edited the content as needed and takes full responsibility for the content of the publication.
\bibliographystyle{JHEP}
\bibliography{main} 
\end{document}